\documentclass[12pt]{article}

\usepackage{soul}

\usepackage{tikz}
\usepackage{stmaryrd}
\usepackage{epsfig,latexsym,amsfonts,amsmath,amsthm,amssymb,amsbsy,multirow,slashed,wasysym,textcomp,subfigure,wrapfig,datetime,comment,mathtools,cancel,cite,mathrsfs}
\usepackage[hidelinks]{hyperref}
\usepackage[font={footnotesize},bf]{caption}

\usepackage{soul}
\usepackage[normalem]{ulem}

\def\ee{\end{equation}}
\def\be{\begin{equation}}
\def\bea{\begin{eqnarray}}
\def\eea{\end{eqnarray}}
\newcommand{\beq}{\begin{eqnarray}}
\newcommand{\eqq}{\end{eqnarray}}
 \newcommand{\badat}{\begin{alignedat}}
 \newcommand{\eadat}{\end{alignedat}}

\newcommand{\eal}[1]{\be \begin{aligned} #1 \end{aligned}\end{equation}} 
\newcommand{\eqn}[1]{\be #1 \end{equation}} 
\newcommand{\eqa}[1]{\bea  #1\end{eqnarray}}

\renewcommand{\d}{\mathrm{d}}

\def\cO{{\cal O}}
\long\def\new#1\endnew{{\bf #1}}		
\long\def\del#1\enddel{}

\def\o{\omega}

\def\cred{ \color{red}}

\def \hp{ \hat \Phi}
\def\cl{{\rm L}}
\def\rj{{\rm J}}

\usepackage{color}

\newcommand{\pink}[1]{\textcolor{\pink}{#1}}

\definecolor{dblue}{rgb}{0.2,0.50,0.80}

\def\D{{\Delta}}

\def\o{\omega}
\def\G{{\Gamma}}

\newcommand{\br}[1]{\overline{#1}}

\def\cL{{\cal L}}

\def\bz{{\bar z}}
\def\bw{{\bar w}}

\def\co{{\cal O}}
\def\scri{{\mathcal{I}}}
\def\d{\delta}
\def\hp{{\bf \Phi}}
 \def\e{\epsilon}

\def\p{\partial}
\def\ec{${\rm{EC}}^4$}
\def\m{${\rm{M}}^4$}
\def\ads{${\rm{AdS}}_4$}
\def\ms{{ T}}
\def\half{{1\over 2}}
\begin{document}
\begin{titlepage}
\unitlength = 1mm~\\
\vskip 3cm
\begin{center}
\def\bl{\big[   }
\def\br{\big]  }
\def\im{{mathcal{I}_-}}
\def\cred{\color{red}}
\def\bigl[{\bigl[}
\def\big\rrbracket{\bigr]}
{\LARGE{Soft Algebras in \ads\ from \\
\vspace{.6cm}
Light Ray Operators in CFT$_3$ }}

\vspace{1cm}Ahmed Sheta, 
Andrew Strominger, Adam Tropper and Hongji Wei \\
\vspace{.7cm}

{\it  Center for the Fundamental Laws of Nature,\\ Harvard University, Cambridge, MA USA } \\

\vspace{0.8cm}

\begin{abstract}

Flat Minkowski space (M$^4$)  and \ads\ can both be conformally mapped to the Einstein cylinder. The maps may be judiciously chosen so that some null generators of the $\scri^+$ boundary of  M$^4$  coincide with antipodally-terminating null geodesic segments on the boundary of \ads. Conformally invariant nonabelian gauge theories in \m\ have an asymptotic  $S$-algebra  generated by a tower of soft gluons given by weighted  null line integrals on  $\scri^+$. We show that, under the conformal map to \ads, the leading soft gluons are dual to  light transforms of the conserved global symmetry currents in the boundary CFT$_3$. The tower of light ray operators obtained from the $SO(3,2)$ descendants of this light transform  realize a full  set of generators of the $S$-algebra in the boundary CFT$_3$. This provides a direct connection between holographic symmetry algebras in \m\ and \ads. 
 \end{abstract}

\end{center}

\end{titlepage}

\tableofcontents
\section{Introduction}
\subsection{Summary}
   Quantum gravity and gauge theory in four-dimensional asymptotically flat spacetimes exhibit infinite-dimensional symmetry algebras variously referred to as 
   soft algebras or celestial chiral algebras.
In quantum field theory or gravity, these arise as the algebra of a tower of soft theorems \cite{Guevara:2021abz, Strominger:2021mtt}, in twistor theory as a symmetry of the Penrose-Ward construction\cite{Penrose:1976jq,Ward:1977ta}, and in twisted holography \cite{Costello:2022wso,Costello:2022jpg} as a generalized 2D chiral algebra on the celestial sphere. The emergence of identical  algebras\footnote{Soft $theorems$ are less universal than soft $algebras$ and typically  receive corrections, especially at subleading orders, while the commutator algebra of successive soft insertions sometimes remains undeformed.} within these disparate approaches indicates their universality. A unified framework relating these approaches was given for gravity in \cite{Adamo:2021lrv} and for gauge theory in \cite{Kmec:2025ftx}.

   Soft algebras are of central interest in the search for a holographic dual to quantum gravity in asymptotically flat spacetimes for the simple reason that both members of a dual pair must exhibit the same symmetries. Invariance under the infinite-dimensional soft symmetry algebra greatly constrains the possibilities for a boundary dual. Indeed, many structural  aspects of  AdS$_3$/CFT$_2$ duality follow from  an infinite-dimensional conformal symmetry group \cite{Brown:1986nw}, independently of its stringy realizations  \cite{Maldacena:1997re,Aharony:1999ti}.

   One might expect realizations of the holographic principle \cite{tHooft:1993dmi,Susskind:1994vu} in different spacetimes --- we have in mind here \ads\ and Minkowski space (\m) --- to carry a common thread. From this point of view it is perhaps surprising that soft algebras have not so far been directly identified in \ads\ for nonzero $\Lambda$.\footnote{They do arise in an appropriately defined limit $\Lambda \to 0$ \cite{deGioia:2023cbd,deGioia:2024yne,deGioia:2025mwt}.} Further suspicion of the existence of \ads\ soft algebras comes from the observation of
   Ward \cite{Ward:cc} that both self-dual gauge fields and Einstein metrics in spaces with negative cosmological constant are in one-to-one correspondence with those in flat space.\footnote{References \cite{Ward:cc,Taylor:2023ajd, Bittleston:2024rqe} do not impose boundary conditions of the type employed in AdS/CFT. The main consequences of these boundary conditions is a restriction to linearly polarized gluons which are shown in sections 8 and 10 to generate an isomorphic $S$-algebra.} In the gauge theory case conformal symmetry implies  they are generated by the same `$S$-algebra' as in flat space. In the gravity case, which is not conformally invariant, one encounters the very interesting  `deformed $w$-algebra'\cite{Taylor:2023ajd, Bittleston:2024rqe}. All of this suggests the possibility  that   soft algebras might provide a unifying theme between  differing realizations of  holography. 

Indeed we will show in this paper that the nonabelian soft gauge algebras of \m\ conformally map to the commutator  algebra of  light transforms   of  conserved currents  (and conformal descendants thereof) in the boundary CFT$_3$ of \ads.  The leading soft generator in \m\ can be realized as the  integral of the gluon field strength along a  null geodesic  in $\scri^+$. We show herein that the remaining tower of $S$-generators can be constructed as $SO(4,2)$ conformal descendants. \ads\ and \m\ are related via the Einstein cylinder (\ec) by conformal mappings. Judiciously chosen conformal maps are constructed with  the property that some of the null geodesics of $\scri^+$ are mapped to null geodesic segments in the boundary of \ads. Using the \ads/CFT$_3$ bulk-to-boundary dictionary \cite{Aharony:1999ti,Aharony:2008ug}, the integrals of 
   gluon field strengths are then mapped to light transforms of  conserved global  currents in the dual boundary CFT$_3$. Their 3D commutator algebra is the subalgebra of the $S$-algebra arising from the leading soft theorem. The $SO(3,2)$ descendants of these current   light transforms are then shown to  fill out  a complete set of generators of the $S$-algebra.

   In summary, the soft $S$-algebra of nonabelian gauge theory in \m\ conformally maps to  the algebra of  light transformed conserved currents and their $SO(3,2)$ descendants in the dual CFT$_3$ on the boundary of \ads. 
   Hence, unconfined nonabelian gauge theories lead to the same soft algebra in both \m\ and \ads.\footnote{Our analysis here does not directly suggest any relation between \ads\ and \m\ holography beyond the soft sector.} One hopes that this may enable the import of  ideas from \ads\ holography, as well as relevant results on CFT light ray operators \cite{Hofman:2008ar,Casini:2017roe,Karateev:2017jgd,Cordova:2018ygx,Kravchuk:2018htv,Kologlu:2019mfz,Kologlu:2019bco,Belin:2020lsr,Huang:2019fog,Huang:2020ycs,Kabat:2021akg},  towards a better understanding of flat space holography.\footnote{ The non-locality of light ray operators is reminiscent of various non-localites encountered in celestial holography, and may shed light on the latter.} 

    We anticipate   a similar result for gravity, with the deformed $w$-algebra generated by an $SO(3,2)$ multiplet of light ray operators including the ANEC operator. The lack of conformal invariance in gravity may be responsible for the $\Lambda$-deformation of the soft algebra. We also anticipate that the methods of this paper can be extended to theories which are not exactly conformally invariant, but flow to a nontrivial conformally invariant fixed point in the IR. In this case Wilsonian corrections will in general deform the soft algebra. For example minimal coupling to gravity mixes the soft $S$- and $w$-algebras. 

   The presence of a soft algebra in \ads\ may be surprising in view of the energy gap which would seem to preclude a soft limit.  Importantly, the term  `soft' here for both \m\ and \ads\ refers to boost weight rather than global energy, which is not gapped.

A number of papers, beginning with \cite{Cordova:2018ygx}, have discussed the relation between 4D (rather than 3D) light ray operators and 4D asymptotic symmetry algebras in a variety of contexts\cite{Mishra_2018,Mishra_2019,Kologlu:2019mfz,He:2019ywq,Donnay:2020fof,Gonzo:2020xza,Hu:2022txx,Donnay:2022aba, Donnay:2022wvx,Hu:2023geb,Gonzalez:2025ene,Moult:2025njc,Himwich:2025ekg}. This 4D-4D connection arises because the asymptotic charges can be written as a sum of a hard and a soft part, which are then related by charge conservation. The hard charges are in some cases given by light ray operators in the bulk 4D theory. In the current work, 4D soft charges are holographically dual  to  3D light ray operators in the boundary CFT$_3$.  
   
\subsection{Outline and discussion}

We begin in section  2 by defining the leading  soft gluon operator  $S^{1,a}_{0,m}$ in \m\ as an integral of the field strength over a null generator of $\scri^+$. We present the leading `$p=1$' subalgebra of the $S$-algebra which follows from the leading soft theorem. Section 3  presents the 15 conformal Killing vectors that generate the $SO(4,2)$ conformal symmetry of \m. In section 4 we show that the $SO(4,2)$ descendants of the leading $S^{1,a}_{0,m}$ generators populate the full set of $S$-generators denoted $S^{p,a}_{\bar m,m}$. Moreover we show by construction that they generate the complete  $S$-algebra. Section 5 shows explicitly that the gluon wavefunctions derived from soft theorems agree with those obtained by a conformal transformation, up to boundary terms encountered in integrating by parts on $\scri^+$. In many cases these boundary terms do not contribute but, wherever they do matter, we regard the definition of $S^{p,a}_{\bar m,m}$ as conformal descendants as more fundamental. 

These sections establish that any conformal theory with an undeformed, leading nonabelian soft theorem has the full $S$-algebra. There are many examples of such theories at the classical level, but quantum examples are rare.\footnote{ The light-ray $S$-algebras appearing in the conformal map to \ads\ are potentially well-defined at the quantum level.  } 
In the pure glue quantum theory, for example, the running of the coupling constant deforms the leading soft theorem. There are no asymptotic gluons, let alone an $S$-algebra.\footnote{Although the existence of a broken $S$-algebra may still be of interest.} One relevant example is the  self-dual Yang-Mills theory as defined in \cite{Chalmers:1996rq}, which has  an exact leading soft theorem, conformal invariance and a nontrivial one-loop-exact $\cal S$-matrix. The existence of an $S$-algebra in the quantum theory was in fact already verified in \cite{Ball:2021tmb} by analysis of the splitting functions. A very rich set of self-dual examples has been discovered among  the twistorial quantum field theories \cite{Costello:2018zrm,Costello:2022wso,Costello:2022jpg,Costello:2023vyy,Bittleston:2024efo}. These all have vanishing $\cal S$-matrices but the $S$-algebra is nevertheless realized as a chiral algebra on the celestial sphere. 

An important example {\it might} be provided by $\mathcal{N}=4$ Yang-Mills in the interacting nonabelian Coulomb phase.\footnote{Others are the finite $\mathcal{N}=2$ finite theories, the Banks-Zaks model \cite{Banks:1981nn} and the $\mathcal{N}=1$ Seiberg models \cite{Seiberg:1994pq}.  }
Unfortunately the $S$-algebra is usually described as acting on the $\cal S$-matrix, which is not defined in $\mathcal{N}=4$ due to IR divergences.  It would be very interesting to know if it acts on suitable IR finite observables in this theory. 
   
Section 6 details the geometry of the conformal mappings between \ads, \ec\  and \m. The light cone of a point ``$i^0$" in \ec\ tessellates it into a series of diamonds, each of which is conformal to flat \m. \ec\ is also conformal to two copies of \ads\ separated by an $S^2\times R$ boundary denoted $\p$\ads. We choose $i^0$ to lie in $\p$\ads, so that some null generators of $\scri^+$ of \m\ are also in $\p$\ads. 
In section 7 we use these conformal mappings to  construct the global extension of conformal scalar and gluon modes on \m\ to \ec.

Section 8 extends  the leading soft gluon wavefunction in \m\ to \ads. Here the \ads\ boundary conditions become important.  In \m\ we usually describe the $S$-algebra as an algebra of positive helicity soft gluons.  However, the usual \ads\ boundary conditions\footnote{CPT-violating boundary conditions allow a single helicity, see the discussion in section 8.} flip helicity and do not allow a positive helicity operator alone at the boundary. Nevertheless we show that the boundary conditions do allow  a diagonal subgroup of linearly polarized soft gluons with an isomorphic leading soft algebra. 

In Section 9 we interpret this leading soft gluon as a boundary operator in the dual CFT$_3$. A bulk gauge field in AdS$_4$ with suitable boundary conditions implies a conserved global symmetry current $\rj^a_i $ in the boundary CFT$_3$. The leading soft gluon is then identified as the light transform of $\rj^a_i $ along a null geodesic beginning at the `north pole' in $\p$\ads\ and ending at the `south pole'.  We show its known commutators as computed in \cite{Cordova:2018ygx} generate the leading $S$-algebra.

   Finally in section 10, mimicking the \m\ analysis of section 4, we use the  $SO(3,2)$  symmetry of \ads/CFT$_3$ to construct descendants of the leading soft/light ray  operator. These are  shown to generate the full $S$-algebra.

\section{Leading soft generator }
In this section we recap the construction of the $S$-algebra in \m.

Let 
$v$ ($u$) be advanced (retarded) time
and $(z,\bz)$ complex coordinates on the celestial sphere 
given in  Cartesian coordinates $X^\mu$ on \m\ by
\be \label{zbz} z={X^1+iX^2 \over X^0+X^3},~~~~v=X^0+X^3,~~~~u+ v z \bar z=X^0-X^3.\ee
The flat metric is 
\be\label{fm} ds^2=-dudv+v^2dzd\bz. \ee
Here $u$ is a null coordinate on $\scri$ and 
$\scri^\pm$ is located at $v=\pm\infty$.
The highest-weight  entry in the tower of generators of the (outgoing) $S$-algebra is  an integral over a null generator of $\scri^+$ at $v=\infty$\footnote{We use the convention $\mathcal{L}= -\frac{1}{4g_{YM}^2}F^a_{\mu\nu}F^{a\mu\nu}$ with $F^a_{\mu\nu} =\partial_\mu A_\nu^a - \partial_\nu A_\mu^a + f^{abc}A^b_\mu A^c_\nu$ and set $g_{YM}=1$.  } \cite{Strominger:2013lka,He:2015zea,Strominger:2017zoo}
\be\label{sonef} S^{1,a}(z)= \sum_m{S^{1,a}_{0,m}\over z^{m+1}}=-4\pi\int_{\mathcal{I}^+} du F^a_{uz}(u,+\infty,  z,\bz) .\ee
where $F^a_{uz}$ is a self-dual field strength operator with adjoint index $a$  which creates an outgoing positive helicity soft gluon. The leading soft theorem  implies that amplitudes involving these soft gluons are  holomorphic in $z$ and have the OPE, for any local charged field in the adjoint  
\be S^{1,a}(z)X^{b}(\bw, w)\sim-{i\over z-w}f^{abc}X^{c}(\bw, w).\ee
Taking $X^b=S^{1,b}$ itself implies the radial 2D commutator 
\be\label{ssone} [S^{1,a}_{0,m},S^{1,b}_{0,n}]=-if^{abc}S^{1,c}_{0,m+n}.\ee 
This 2D commutator is defined for holomorphic fields 
\be X^I(z)=\sum_m {X^I_m\over z^{m+h_I}} \ee
by
\be [X^I_m,X^J_n]=\oint_0 \frac{dw}{2\pi i} w^{n+h_J-1}\oint_w \frac{dz}{2\pi i} z^{m+h_I-1}X^I(z)X^J(w).\ee
It picks out poles in the $X^I(z)X^J(w)$ OPE.

The $SO(3,1)$  Lorentz group has generators $L_n, \bar L_n$ for $n=0,\pm 1$. $S^{1,a}$ transforms as a holomorphic current
\be \Bigl[\bar L_{\bar n},S^{1,a}_{0,m}\Bigr]  =0,~~~~\Bigl[  L_n,S^{1,a}_{0,m}\Bigr]  =-mS^{1,a}_{0,m+n}.\ee
Here the large bracket denotes the standard 4D commutator. Lorentz invariance implies  that the 4D commutator defines an outer derivation of the 2D commutator
\be\label{assc} \Bigl[  L,[S,S']\Bigr] + [ S',\Bigl[ L,S\Bigr] ]+[ S,\Bigl[ S',L\Bigr] ]=0.\ee

\section{$SO(4,2)$ conformal transformations}

The OPEs of soft gluons are governed by  the 4D  conformal symmetry arising from  15 conformal Killing vectors (CKVs) whose  Lie bracket algebra is the  $SO(4,2)$ Lie algebra. For the flat metric in coordinates \eqref{fm} these are 
dilations
\be 
 D = u\,\partial_u + v\,\partial_v,\ee
Lorentz transformations 
\bea &&L_{-1} = -\partial_z,\quad L_0 = -z\,\partial_z-\tfrac12 u\,\partial_u+\tfrac12 v\,\partial_v,\quad L_{1} = -z^2\,\partial_z-z \,u \,\partial_u + z\,v\,\partial_v \, + \frac{u}{v}\partial_{\bar{z}},\cr
&& \bar L_{-1} = -\partial_{\bar z},\quad \bar L_0 = -\bar z\,\partial_{\bar z}-\tfrac12 u\,\partial_u+\tfrac12 v\,\partial_v,\quad \bar L_{1} = -\bar z^2\,\partial_{\bar z} -\bar z\, u \,\partial_u+ \bar z\,v\,\partial_v+ \frac{u}{v}\partial_z,\eea
translations
\bea &&P_{-\frac{1}{2},-\frac{1}{2}}  =  -i\partial_u\cr &&
P_{-\frac{1}{2},\frac{1}{2}}\hspace{7pt} = -i(z\,\partial_u - \tfrac{1}{v}\,\partial_{\bar z})\cr &&
P_{\frac{1}{2},-\frac{1}{2}} \hspace{7pt} = -i(\bar z\,\partial_u - \tfrac{1}{v}\,\partial_{z})\cr &&
P_{\frac{1}{2},\frac{1}{2}} \hspace{14pt} = -i(z\bar z\,\partial_u - \tfrac{1}{v}(z\,\partial_z + \bar z\,\partial_{\bar z}) + \,\partial_v),\eea
and special conformal transformations

\bea\label{kckv} && K_{-\frac{1}{2},-\frac{1}{2}}  = iv^2\,\partial_v\cr &&
K_{-\frac{1}{2},\frac{1}{2}} \hspace{7pt} =  i(z\,v^2\,\partial_v + u\,\partial_{\bar z})\cr &&
K_{\frac{1}{2},-\frac{1}{2}} \hspace{7pt}= i(\bar z\,v^2\,\partial_v + u\,\partial_{z})\cr &&
K_{\frac{1}{2},\frac{1}{2}} \hspace{14pt} = i(z\bar z\,v^2\,\partial_v + u(z\,\partial_z + \bar z\,\partial_{\bar z}) + u^2\,\partial_u) .\eea
 
In the quantum theory these symmetries are generated by 4D commutators with the associated Noether charge which, at risk of confusion, we denote by the same symbol. We normalize these charges so that for a scalar primary field of weight $\chi$
\begin{equation}
     \quad  \Bigl[ Q_\zeta, \co\Bigr] = -(\cL_\zeta +\frac{\chi}{4} \nabla \cdot \zeta ) \mathcal{O}
\end{equation}
where $\zeta$ is any one of the 15 CKVs and $\Bigl[Q_{\zeta_1},Q_{\zeta_2}\Bigr]=Q_{[\zeta_1,\zeta_2]}$. The nonzero charge commutators are 
\begin{equation}    \begin{split}
        \Bigl[  L_n,L_m\Bigr]  &= (n-m)L_{n+m} \hspace{32pt} \Bigl[  \bar {L}_{\bar n},\bar{L}_{\bar m}\Bigr]   = (\bar n-\bar m) \bar{L}_{\bar n+\bar m} \\
        \Bigl[  L_n,P_{\bar{r},r}\Bigr]   &= \tfrac{1}{2}(n-2 \hspace{1pt}r)P_{\bar{r},r +n} \hspace{18pt} \Bigl[  \bar{L}_{\bar n},P_{\bar{r},r}\Bigr]   = \tfrac{1}{2}(\bar n-2\hspace{1pt}\bar{r})P_{\bar{r} + \bar n,r} \hspace{16pt} \Bigl[  D,P_{\bar{r},r}\Bigr]   = -P_{\bar{r},r }\\
        \Bigl[  L_n,K_{\bar{r},r }\Bigr]   &= \tfrac{1}{2}(n-2 \hspace{1pt}r)K_{\bar{r},r +n} \hspace{13pt} \Bigl[  \bar{L}_{\bar n},K_{\bar{r},r}\Bigr]   = \tfrac{1}{2}(\bar n-2\hspace{1pt}\bar{r})K_{\bar{r}+\bar n,r} \hspace{11pt} \Bigl[  D,K_{\bar{r},r}\Bigr]   = K_{\bar{r},r}\\
        \Bigl[  K_{\bar{r},r },P_{\bar{s},s}\Bigr]   &= {-\epsilon_{\bar r, \bar s} \epsilon_{r, s} D - \epsilon_{\bar r, \bar s} L_{r+s} - \epsilon_{r,s} \bar L_{\bar r+\bar s}}
        \label{eqn: so(4,2) commutation}
    \end{split}
\end{equation}
where $\epsilon_{-\frac{1}{2},\frac{1}{2}} = -\epsilon_{\frac{1}{2},-\frac{1}{2}} = 1$.
\section{Full $S$-algebra from conformal descendants}
\label{sec: S algebra from conf descendents}
States and operators in a theory with  conformal and  leading soft symmetries must fall into  representations of the associated algebras. However, the leading soft generators  do not themselves  fill out a representation of $SO(4,2)$. Indeed, it was shown already in \cite{Larkoski:2014hta} that the subleading soft theorem in Yang-Mills theory is related by a special conformal transformation to the leading one. 
In this section we define the full set of generators $S_{\bar m,m}^{p,a}$ using the action of $SO(4,2)$ on the leading generators $S^{1,a}_{0,m}$. The equivalence of this (up to boundary terms) with previous integral definitions of $S_{\bar m,m}^{p,a}$ will be demonstrated in the next section.  

The leading $p=1$ generators comprise a representation  of the 2D Euclidean conformal group $SO(3, 1)$ but not of $SO(4,2)$.  Elements of an  $SO(4,2)$ representation are labeled by  the eigenvalues of the Cartan subalgebra generated by  $D$, $L_0$ and $\bar L_0$, or equivalently $p$, $m$ and $\bar m$ where $2(p-1)$ is the eigenvalue of $D$. $S_{0,m}^{1,a}$ is dilation invariant, and $K$ and $P$ are raising and lowering operators for $p$.  Here we $\it define$  $S_{\bar m,m}^{p,a}$ for $p={3\over 2}, 2,{5\over 2},\ldots$ and $m\ge {p-1}$  iteratively from $S^{1,a}_{0,m}$ by 
\be \label{eq:[K,s]} \hspace{35pt}\Bigl[  K_{\bar r, r}, S_{\bar m,m}^{p,a}\Bigr]   = (-)^{r-\bar r} ((p-1) +2rm ) S^{p+{1\over 2}, a}_{\bar m + \bar r, m+r} .\ee
\begin{equation}\label{pst}
        \Bigl[   P_{\bar{r},r},S^{p,a}_{\bar{m},m} \Bigr]  = ((p-1)- 2\bar{r} \bar{m}) S^{p-1/2,a}_{\bar{m} + \bar{r},m+r} 
        \ee
        \be\label{dst}
        \hspace{-52pt}\Bigl[ D,S^{p,a}_{\bar{m},m}\Bigr]   = 2(p-1) S^{p,a}_{\bar{m},m}
   \end{equation}\label{bnm}
   \be \Bigl[   L_n,S^{p,a}_{\bar{m},m}\Bigr]  = (n(1-p)-m)S^{p,a}_{\bar{m},m+n} ,~~~~\Bigl[  \bar L_{\bar{n}},S^{p,a}_{\bar{m},m}\Bigr]  =(-)^{\bar{n}}(\bar{n}(p-1)-\bar{m})S^{p,a}_{\bar{m}+\bar{n},m}. \label{l commutator}\ee
The coefficients here are consistent with the $SO(4,2)$ commutators \eqref{eqn: so(4,2) commutation}. Indeed they are unique up to renormalizations of  $S_{\bar m,m}^{p,a}$, which by construction are in a representation of the $SO(4,2)$ algebra. The representation is lowest weight with respect to $P_{\bar{r},r}$ and has vanishing Casimirs because all elements of $SO(4,2)$ annihilate $S^{1,a}_{0,0}$.

We expect soft generators for every value of  $p=1,{3\over 2}, 2,{5\over 2},\ldots$, every value of $m+p\in \mathbb{Z}$ and every $\bar m + p \in \mathbb{Z}$  in the $\bar m$-wedge   $|\bar m| < p$.
It is easy to see that, starting from $S^{1,a}_{0,m}$,  all such soft generators can be reached $except$ those in the $m$-wedge $|m| <{p-1}$. Entry into this wedge by $SO(4,2)$ action is prevented by  zeroes in the prefactors in \eqref{eq:[K,s]}. However, this barrier  is easy to get around.  The leading soft theorem applies to all local charged operators and in particular implies 
\be\label{ssp} [S^{1,a}_{0,m},S^{p,b}_{\bar n,n}]=-if^{abc}S^{p,c}_{\bar n,m+n}.\ee
Using this we can raise or lower the $L_0$ eigenvalue at will, and the combination of conformal and leading soft generators can therefore be used to define the full tower of soft generators. Moreover it may be shown that \eqref{eq:[K,s]}-\eqref{ssp} together define a unique operator for each value of $p, \bar m,  m$. 

The commutators of the $S^{p,a}_{\bar m ,m}$ among themselves are now implied by their definitions and the Jacobi identity. One finds
\be\label{ssxpp} [S^{p,a}_{\bar m ,m},S^{q,b}_{\bar n,n}]=-if^{abc}S^{p+q-1,c}_{\bar m +\bar n,m+n},\ee  which is  the full $S$-algebra.

We conclude that any theory that has $SO(4,2)$ invariance plus the  leading soft theorem \eqref{ssp}  implies the  full $S$-algebra  \eqref{ssxpp}.

\section{Conformal descendants and  Mellin transforms}
 In the previous section the tower of $S$-generators were defined as conformal descendants of the leading $S$-generator and commutators thereof.  A more familiar definition is as residues of poles in the  Mellin transform of an asymptotic  positive helicity gluon. In this section we  show that these definitions agree, up to boundary terms in integration by parts which are not  always specified in various representations of the asymptotic integrals. In many contexts these boundary terms do not contribute. If they do  they may be fixed by the  fundamental definition of the subleading $S$-generators as conformal descendants given in the previous section. 

Let us define the  Mellin transformed asymptotic gluon operator
\bea \label{xdf} &&\hspace{-52pt}X_z^{p,a}(z,\bz) \equiv \hspace{2pt} \frac{{i^{2p-2}}N_p}{\Gamma(2-2p)} \int_0^\infty d\o \hspace{2pt} \omega^{1-2p}\widetilde F^a_{uz}\cr && =N_p\int_{\mathcal{I}^+} du \hspace{2pt} u^{2p-2}F^a_{uz},~~~~~~~~ ~~~N_p ={{4\pi}\hspace{2pt} i^{2p} \over \G(2p-1)}. \eea
where $\widetilde{F}_{uz}(\omega,z,\bar{z}) = \int_{-\infty}^\infty du e^{i u \omega} F_{uz}(u,z,\bar z)$.
For $2p-1 \in {\mathbb{Z}_+}$ the expansion in soft gluon operators is \cite{Strominger:2021mtt} 
\be\label{dse} X_z^{p,a} = \sum_{\bar m=1-p}^{p-1}\sum_{n+p\in \mathbb{Z}}\frac{S^{p,a}_{\bar 
m,n}}{\Gamma(p+\bar m)\Gamma(p-\bar m)} \frac{1}{z^{2-p+n} \bar z^{1-p+\bar m}},\ee
 We will show that a Lie derivative with respect to a CKV on the LHS agrees with the action of the associated $SO(4,2)$ action \eqref{eq:[K,s]} on the modes on the RHS, thereby verifying agreement between Mellin-transformed subleading soft gluons and conformal descendants of the leading soft gluons.  

It follows from \eqref{xdf} that
\be\label{dwx} \Bigl[  D, X_z^{p,a}\Bigr] =2(p-1)X_z^{p,a}.\ee
Since $K$ is a raising operator for $D$ a recursive  relation between $X_z^{p,a}$ and $X_z^{p+\half,a}$ can be obtained from a special conformal transformation
\be\label{dkf} \Bigl[  K_{-\half,-\half}, X_z^{p,a}\Bigr]  = -N_p\int_{\mathcal{I}^+} duu^{2p-2}\cL_{K_{-\frac{1}{2},-\frac{1}{2}}}F^a_{uz}=-i N_p\int_{\mathcal{I}^+} du u^{2p-2}v^2\p_v F^a_{uz},\ee
where the Lie derivative is with respect to the first vector field in \eqref{kckv}.
Combinations of the (linearized) Bianchi and constraint equations imply
\be \p_vF^a_{uz}=-{1 \over v^2}\p_zF^{+a}_{z\bz},~~~~  \p_\bz F^a_{uz}= -\p_u F^{+a}_{z\bz},\ee
where 
$F^{+a}_{z\bz}\equiv \frac{1}{2}(F^{a}_{z\bz}-v^2 F^{a}_{uv})
$ is the self dual part of  $F^{a}_{z\bz}$. Differentiating this identity gives
\be \p_u\p_vF^a_{uz}={1 \over v^2}\p_z\p_\bz F^{a}_{uz}.\ee Integrating by parts with respect to $u$,
\eqref{dkf} then becomes 
\be \label{nji}\Bigl[  K_{-\half,-\half}, X_z^{p,a}\Bigr]  =\p_z \p_\bz X_z^{p+\half,a}.\ee

We may also derive a  recursion relation from the $SO(4,2)$ transformation
 laws for $S^{p,a}_{\bar 
m,n}$ of the previous section. The relevant formula is 
\be \Bigl[  K_{-\half,-\half},S^{p,a}_{\bar m,m}\Bigr] = (p-1-m)S^{p+\half,a}_{\bar m-\half,m-\half}.\ee
This yields\footnote{Note that the top term in the sum over $\bar m'$ is killed by $\p_\bz$.}
\bea   \Bigl[  K_{-\half,-\half}, X_z^{p,a}\Bigr] &=&\sum_{\bar m=1-p}^{p-1}\sum_{m+p\in \mathbb{Z}}\frac{(p-1-m)S^{p+\half,a}_{\bar m-\half,m-\half}}{\Gamma(p+\bar m)\Gamma(p-\bar m)} \frac{1}{z^{2-p+m} \bar z^{1-p+\bar m}}\cr 
&=&\p_z\p_\bz \sum_{\bar m'=1-(p+\half)}^{(p+\half)-1}\sum_{m'+(p+\half)\in \mathbb{Z}}\frac{S^{p+\half,a}_{\bar m',m'}}{\Gamma((p+\half)+\bar m')\Gamma((p+\half)-\bar m')} \frac{1}{z^{2-(p+\half)+m'} \bar z^{1-(p+\half)+\bar m'}}\cr
&=& \p_z\p_\bz X_z^{p+\half,a} .\eea
 in agreement with \eqref{nji} and confirming the expansion  \eqref{dse}.
Inverting we get \begin{equation} \label{eq:S modes}
    S^{p,a}_{\bar{m},n} = N_p \hspace{2pt} \Gamma(p+\bar{m}) \Gamma(p-\bar{m}) \oint \frac{dz}{2\pi i} \frac{d \bar z}{2\pi i} z^{1-p+n} \bar z^{-p+\bar{m}} \int_{\mathcal{I}^+} du\, u^{2p-2} F_{uz}.
\end{equation}

\section{Minkowski (\m), Einstein Cylinder (\ec) and \ads}
\m\ and \ads\ can both be conformally mapped to the $S^3\times \mathbb{R}$ Einstein cylinder (\ec). In this section we describe  a judicious choice of  mapping containing  null line segments in \ec\  living  both in $\scri^+$ and in  the \ads\ boundary. This will enable a direct relation between $S$-algebra  generators in \m\ and light ray operators in the CFT$_3$ on the boundary of \ads. 

\ec\ is conformally equivalent to two copies of \ads\ glued   at a common $ S^2\times \mathbb{R}$ (EC$^3$)
boundary $\p$\ads. Let $i^0$ denote a point in $\p$\ads. The light cone of $i^0$ tessellates $\p$\ads\ into a series of M$^3$ Minkowski diamonds. Let \m\ denote the region of \ec\ which is spacelike separated from $i^0$.  M$^3$ is then a timelike slice of \m. Null infinity $\scri^\pm\sim S^2\times \mathbb{R}$ of 
\m\ are portions of the $i^0$ light cone. 
The null generators of $\scri_2^\pm$(M$^3$)$\sim S^1\times \mathbb{R}$ lie in both $\scri^\pm$ and $\p$\ads.

The choice of a point $i^0$ in \ec\ breaks the SO(4,2) conformal group down to $ISO(3,1)\times D $: Poincare transformations and dilations of \m. The choice of an \ads\ boundary in \ec\ breaks it down to $SO(3,2) $. The intersection of these preserve a common $ISO(2,1)\times D $, the conformal group preserving  M$^3$. 

Both $\scri^+$ and $\scri^-$ are Cauchy surfaces in \ec\ as well as  \m,  when the points $i^0$ and $i^\pm$ are added.  Solutions of a conformally invariant wave equation on \m\ may be continued  to \ec\ and \ads. However if Dirichlet or Neumann boundary conditions are imposed on \ads\  only half the solutions may be so extended.

Soft operators in \m\ may be  expressed as integrals along the null generators of $\scri^+$, as in \eqref{sonef}: 
\be \label{sz} S^{1,a}(z)
= - 4\pi \int_{-\infty}^\infty du F^a_{uz}(z,\bz),\ee
and are parameterized by a point  $(z,\bz)$  on the celestial sphere.  In the conformal compactification to \ec,\footnote{Under this compactification the $SO(4,2)$ invariant vacuum on \m\ maps to that on \ec. However generic states on a spacelike slice of \m\ with non-zero large gauge charges 
$Q[\e(z,\bz)]$ cannot be mapped to smooth states on \ec. We restrict to zero-charge states which can be so mapped, effectively equating the hard and soft parts of $Q[\e(z,\bz)]$ when expressed as integrals over $\scri^\pm$.} these null generators emanate from a single point $i^0$  and hence are characterized by an outgoing angle. A one-parameter family of these geodesics lie in $\p$\ads\ as well as ${\scri}^+$. They reconverge and  end at the antipodal point on $\p$\ads. The map between a point and its antipode is $SO(4,2)$ invariant.

Of the 15 $SO(4,2)$ generators only the $11$ $D, L_n, \bar L_{\bar{n}} , P_{\bar{r},s}$ preserve the \m\ diamond. 
\ads\ on the other hand is preserved by the 10 $SO(3,2)$ generators
\be\label{sott} D,~~~~ L_n-\bar L_{-n},~~~~ K_{\half,-\half}, K_{-\half,\half},(K_{\half,\half}+K_{-\half,-\half}),~~~~
P_{\half,-\half}, P_{-\half,\half},(P_{\half,\half}+P_{-\half,-\half}).\ee While 
some elements of $SO(4,2)$ do not preserve the \m\ diamond, there are no boundary conditions placed at those boundaries, and the operator spectrum, algebra and vacuum obey the full $SO(4,2)$ symmetry. In \ads,
on the other hand, typical boundary conditions explicitly break the symmetry to $SO(3,2)$. The \ads\ vacuum or spectrum are not $SO(4,2)$ invariant.\footnote{The unique  $SO(4,2)$ invariant vacuum state on \ec\ can be described as an entangled state in the pair of \ads\ halves. 
Tracing over the Hilbert space of one member of the \ads\ pair leads to a density matrix in the second given in \cite{Chen:2023tvj}. 
Operator products and correlators in this mixed state will be $SO(4,2)$ invariant. This is consistent with the fact that the \ads\ soft algebra, which should not depend on a choice of state, is $SO(4,2)$ covariant.}

\subsection{Conformal mappings}
This subsection details the conformal mappings relating  \m, \ads\ and \ec.

The metric  on a unit-radius \ec\ is given by 
\be ds^2_{EC_4}=-dt^2 +d\theta^2 +\sin^2\theta (d\psi^2+\sin^2\psi d\phi^2), ~~~\phi \sim \phi+2\pi, ~~~~0\le\theta, \psi\le \pi .\ee
The symmetry group of any conformal field theory on this space is $SO(4,2)$.   The null surface $\cos\theta=\cos t $ (or $t=2\pi n\pm \theta$, $n\in \mathbb{Z}$) divides \ec\ into causal diamonds.  Defining the  Weyl factor{\be \Omega_{M_4}= {1 \over \cos t - \cos \theta },\ee}  one finds that
\be ds_{M^4}^2=\Omega^2_{M^4}ds^2_{EC^4}\ee
in the region $0\le \theta \le \pi, |t|\le \theta$ is the flat  metric on a Minkowski  diamond. 
The past and future celestial spheres are located at $\theta={\pi \over 2}, ~~t=\pm{\pi \over 2}$. 
We shall sometimes employ the complex coordinate\footnote{This $z$ is the curved Bondi coordinate, different from the flat Bondi $z$ defined in \eqref{zbz} and used in the previous sections. The two are equal on $\mathcal{I}^+$.}
\be z=e^{i\phi}\tan {\psi \over 2}\ee
in which the metric on the celestial sphere is  $d\Omega_2^2={4dz d\bz \over (1+z\bz)^2}.$
Cartesian coordinates with  $ds_{M^4}^2=\eta_{\mu\nu}dX^\mu dX^\nu$ are given by 
\be X^{\mu} = {\left (\sin t, \sin \theta \sin \psi \cos \phi,\sin \theta \sin \psi \sin \phi, \sin \theta \cos \psi\right) \over {\cos t-\cos \theta}} \ee
\be= {\left(\sin t, \sin \theta \hat x\right )\over {\cos t-\cos \theta }} , ~~~~~\hat x^2=1.\ee

Now let's split \ec\ in half along the timelike surface 
in\be \p{\rm AdS}_4:~~~\psi ={\pi \over 2}.\ee
Defining the  Weyl factor \be \Omega_{AdS_4}=\frac{1}{\sin \theta \cos \psi} \ee  one finds that
\be \label{adsmt}ds_{AdS_4}^2=\Omega^2_{AdS_4}ds^2_{EC^4}\ee
is the $SO(3,2)$ invariant $AdS_4/\mathbb{Z}$ metric on both sides of the `equator' $\psi={\pi \over 2}$.
Hence \ec\  is the conformal completion of both \ads\ and \m.

At a fixed moment of time, say $t={\pi \over 2}$, spatial sections of \ec\ comprise an $S^3$ in which the boundaries of \ads\ and \m\ are $S^2$ submanifolds. In our embedding these spatial boundaries intersect along the common $S^1$ at their common equator  $\psi={\pi \over 2}$. 
This choice of embeddings of \ads,  \m\ $\in$ \ec\ preserve a common $ISO(2,1)\times D$ subgroup of $SO(4,2)$.\footnote{Had we chosen $i^0$ to be in the interior rather than the boundary of an \ads\ region an $SO(3,1)$ would be preserved.} 

\section{Conformal primary wavefunctions on \ec}
This section presents the extension of the $SO(3,1)$ conformal primary wavefunctions from \m\ to \ec. The subsequent restriction to \ads\ appears in the next section. 
\subsection{Scalars}
Solutions of the massless scalar wave equation can be organized into $SO(3,1)$ Lorentz/ conformal primary wavefunctions paramterized by a unit null vector $q$, or equivalently a point on the celestial sphere, and a conformal weight $\D$. The explicit solutions are \cite{Pasterski:2016qvg} 
\be \label{rma} \phi^\D_{q,\pm}(X) =\int_0^\infty d\omega \omega^{\Delta-1} e^{\pm i \omega q\cdot X -\epsilon \omega} ={\Gamma(\D) ~ i^{\pm \D} \over(q \cdot X\pm i\e)^\D}  , \ee
where 
\be\label{eqn: q EC} q^{\mu} =\left (1,  \sin \psi'  \cos \phi',\sin \psi' \sin \phi',  \cos \psi' \right)\ee
and 
\be q\cdot X= {\sin t-\sin\theta\cos\Omega  \over \cos \theta -\cos t}= {\sin t-\sin\theta \hat q\cdot \hat x  \over \cos \theta -\cos t} \ee
with $\Omega$ the solid angle on $S^2$ between $X$ and $q$. 
We have taken the branch cut so that it equals ${\G(\D) i^{\pm \D} \over|q \cdot X|^\D}$ in the lower region with $q\cdot X>0$, while the phase in the upper region is $i^{\mp \D}$.  This phase prescription makes the modes invariant under $CPT$, defined as $X\to-X$ combined  with complex conjugation.

The operator which creates a standard scalar in conformal primary state in Minkowski space is then made from the symplectic product \cite{Pasterski:2016qvg}
\be\label{cpo} \cO^\D_{q,\pm}=\pm i(\phi^\D_{q,{\mp}}|\hp)_{\Sigma\in M_4},\ee
where $\hp(X)$ is  the local scalar field operator, $\Sigma$ is any complete spacelike slice and \be\label{sp} (\phi_1|\phi_2)_\Sigma\equiv \int_\Sigma d^3\Sigma^\mu \phi_1{\overleftrightarrow \nabla_\mu}\phi_2.\ee Pushing the slice up to the $\scri^+$ boundary of $M_4$, one finds the support of $\phi^\D_{q,\pm}$ localizes to a single null generator and the usual expression for $\cO^\D_{q,\pm}$ as a Mellin transform is recovered. 

We wish to extend these primary fields and operators from \m\ to \ec. Under 
a Weyl transformation    \be g\to\Omega^2 g,~~~\phi^\D_{q,\pm}\to\Omega^{-1} \phi^\D_{q,\pm},~~~\hp \to\Omega^{-1}\hp,~~~ \cO^\D_{q,\pm} \to \cO^\D_{q,\pm}.\ee 
The invariance of $\cO$  follows from the invariance of the symplectic product \eqref{sp}. Extending the wavefunctions 
$\phi^\D_{q,\pm}$ to \ec\ requires care in crossing branch cuts at  $\scri^\pm$. 

One finds that the unique single-valued global extension of \eqref{rma} is 
\be\label{efss}\phi^\D_{q,\pm}= - i^{\pm \Delta} \Gamma(\Delta) {(\cos \theta -\cos t\mp i\e \sin t )^{\D-1}\over( \sin t-\sin\theta \hat q\cdot \hat x \mp i\e\cos t )^\D} .\ee  We note that the surface $\cos\theta=\cos t$ is the light cone of $i^0$ at  $(\theta,t)=(0, 0)$, while 
$\sin t=\sin\theta \hat q\cdot \hat x $ is the light cone of the point $(\theta,t, \hat x)=({\pi \over 2},{\pi \over 2},\hat q)$ denoted $q^0$. Taking $\hat q=(0,0,1)$ these intersect at the north pole $\psi =0$ along the null line $\theta=t$. This is a closed null circle comprising the north pole on $\scri^+$ and the south pole on $\scri^-$.  The intersection is codimension three rather than two because the light cones just `kiss'.
The branch cut prescription is defined by analytic continuation of  $t\to t\mp i\e$.\footnote{For $\hat q=(0,0,1)$  the branch cuts intersect in $CS^2$  at $(t,\theta,\psi, \phi)= ({\pi \over 2},{\pi \over 2}, 0,\phi)$ in $CS^2$. Defining $\theta={\pi \over 2}+y$, $t = {\pi \over 2} + t'$ and expanding around this point one finds
\be \label{efxss}\phi^\D_{q,\pm}(t, \theta, \hat x)\sim  {(y- t'\pm i\e  )^{\D-1}\over( -t'^2+y^2+\psi^2\pm i\e t' )^\D} ,\ee
which is a source-free solution of the laplacian. }  Since $t$ and $X^0$ are both timelike coordinates everywhere in \m, \eqref{efss} agrees therein  with the conformal transformation of  \eqref{rma}. 

Solutions of the conformal wave 
 equation on \ec\  should be periodic under $t\to t+2\pi$. Let us check for consistency. The factor 
$\cos \theta -\cos t\mp i\e \sin t $ acquires a net phase of $e^{\pm 2\pi i}$ under this shift, so the numerator itself acquires a phase of $e^{\pm 2\pi i\D}$ and is not single-valued. However the denominator acquires the same phase and the full expression for $\phi^\D_{q,\pm}$ is well defined. Note that this would not be the case if we choose different $i\e$ prescriptions for the numerator and denominator.  The phase choice implies that the modes are invariant up to a minus sign under the antipodal map comprising the $\mathbb{Z}_2$ center of $SO(4,2)$
\be A(t, \theta, \hat x)=(t+\pi, \pi-\theta, -\hat x)\ee
combined with complex conjugation.

At integral $\D$, the case of interest for the soft modes, no phases are acquired at the singularities. The $\pm$ solutions agree everywhere except for distributional differences along the light cones of $i^0$ and $q$.

\subsection{Gluons}

A conformal primary wave function for a positive helicity gluon is \cite{Pasterski_Shao_2017}
\be A^{\D}_{\mu,q,\pm}=\phi^\D_{q,\pm}\e_\mu, ~~~\e_\mu=\p_wq_\mu.\ee
Here and in the rest of this section all operators are positive helicity and the helicity index  is suppressed. 
A conformal primary operator is obtained from the symplectic  product on a complete $S^3$ slice $\Sigma$:
{\be\label{opre}
    \cO^{\D,a}_{q,\pm}=\pm i(A_{\mp}^\D, \boldsymbol{A}^a)= \pm i \int_\Sigma \left[A_{\mp}^\D\wedge *\boldsymbol{F}^a- \boldsymbol{A}^a\wedge*F_{\mp}^\D\right]
\ee}
where  $*F=iF$ for positive helicity. Pushing $\Sigma$ to $\scri^+$ this reduces to the usual Mellin formula. 

To extend this to \ec\ we first transform to global coordinates
\be \p_wq_\mu dX^\mu= d\left[\frac{\sin \theta\partial_w\hat{q}\cdot \hat{x}}{\cos t - \cos \theta}\right], \hspace{24pt} \ee
Using the fact that there is no Weyl rescaling of $A_\mu$ we have the global extension
\be A^{\D}_{\mu,q,\pm} dX^\mu=\Gamma(\Delta)i^{\pm \Delta}\Bigr({\cos \theta -\cos t\mp i\e \sin t \over \sin t-\sin\theta \hat q\cdot \hat x \mp i\e\cos t }\Bigl)^\D d\left[\frac{\sin \theta\partial_w\hat{q}\cdot\hat{x}}{\cos t - \cos \theta}\right].\ee
The tower of soft gluon operators are then
{\be
X^{p,a}_\pm(w,\bar{w}) = {N_p\over 4\pi}   \frac{(\pm i)^{2p}}{(1+w \bar{w})^{2-2p}\Gamma(2-2p)}\int_\Sigma \left[A_{\mp}^{3-2p}\wedge *\boldsymbol{F}^a- \boldsymbol{A}^a\wedge*F_{\mp}^{3-2p}\right]
\ee}
with $w={q^1+iq^2 \over q^0+q^3}$.\footnote{The $1+w\bar{w}$ factor results from our convention \eqref{eqn: q EC}, and can be eliminated by rescaling $q_\mu$.}

\subsection{Gauge invariance?}

One might conclude from the preceding that we have constructed an $S$-algebra on \ec. But it is not so simple. The problem is that the operator \eqref{opre} is not gauge invariant and so is ill-defined outside of perturbation theory in a fixed gauge. This difficulty is circumvented in \m\ when $\Sigma$ is pushed to the \m\ boundary $\scri^+$, where the color frame is fixed  in the computation of scattering amplitudes.\footnote{Of course, as discussed in the introduction, the issue typically reappears as IR ambiguities of the $\cal S$-matrix at the loop level.} While there may be some way to define an $S$-algebra in \ec, \eqref{opre} is insufficient. 

Our main interest in this paper is \ads, for which \ec\ is only an intermediate step. In that case,  boundary conditions can also fix  a color frame at the boundary.  In the next two sections we will see how to lift the \m\ $S$-algebra to \ads\ after properly accounting for the  boundary conditions.

\section{Leading soft operators in \ads}
\label{sec: leading soft operator in AdS}
The operator $S^{1,a}(z)$ does not quite act as a  boundary operator in \ads\ because it does not obey the usual boundary conditions such as 
\be\label{ubc} n\wedge F^a|_{\p\rm{AdS}_4 }=0\ee
where $n=d\psi$ is the normal to boundary. This Dirichlet condition reflects a positive to a negative helicity  gluon and so is inconsistent with self-duality. \eqref{ubc} is  modified by the addition of a theta term  $-\frac{\theta}{16\pi^2}\int F^a\wedge F^a$ to the action, but no real value of $\theta$ allows the self-dual operator $S^{1,a}(z)$ at the boundary. For the special CPT-violating $imaginary$ value $\theta={8\pi^2i  \over  g^2}$ self-dual excitations are allowed. Very interesting recent progress in understanding this case appears in \cite{ Jain:2024bza, Aharony:2024nqs}. 

In this paper, however, we shall consider conventional boundary conditions of the form \eqref{ubc}, and show that the linear combinations of $S^{1,a}$ and its Hermitian conjugate $\bar S^{1,a}$ allowed at the boundary obey a $p=1$ $S$-subalgebra. For these purposes it is convenient to use cylindrical coordinates on the celestial sphere
\be w=-i\ln z=\phi  -i\ln{\tan{\psi \over 2}}\sim w+2\pi,\ee for which 
the \ads \ boundary is on the real axis at \be w=\bw \ee and the normal is $n={i\over 2}(dw-d \bar{w})$.
 \eqref{ubc} allows boundary operators constructed from 
\be  F_{uw} - F_{u\bw}. ~~~~\ee  These have polarization vectors normal to the \ads\ boundary and tangent to the celestial sphere. The  linear combination
\be \label{sdef}\ms^{1,a}(w, \bw)=\frac{1}{2 i} \big(S^{1,a}(w)- \bar S^{1,a}(\bw)\big), \ee
is then allowed for $w=\bar w=\phi$ at the boundary. Moreover it can be seen from formulae below that it transforms as an $SO(3,2)$ primary operator. 
Mode expanding   on the real axis in the convenient  $SO(4,2)$ covariant basis this becomes 
\be \label{sfdef}\ms^{1,a}_m= \frac{1}{2}\big(S^{1,a}_{0,m}+\bar S^{1,a}_{-\bar{m},0}\big). \ee  This is the familiar  projection of a left-right current algebra in a CFT$_2$ in the presence of a boundary:   $\p$\ads\ slices the CCFT$_2$ on the celestial sphere in half and inserts a boundary condition at the equator.  $\bar S^{1,a}_{-\bar{m},0} $ in   \eqref{sfdef} here is the mode of the negative helicity gluons whose $SO(4,2)$ transformation laws are 
\begin{equation}
    \begin{split}\Bigl[  K_{\bar r r}, \bar S_{\bar m,m}^{p,a}\Bigr]   &= (-)^{r-\bar r} ((p-1)+2 \bar r \bar m) \bar S^{p+{1\over 2}, a}_{\bar m + \bar r, m+r},~~~~
        \Bigl[   P_{\bar{r}r},\bar{S}^{p,a}_{\bar{m},m} \Bigr]  = ((p-1)- 2 r m) \bar{S}^{p-1/2,a}_{\bar{m} + \bar{r},m+r} \\
        \Bigl[ D,\bar{S}^{p,a}_{\bar{m},m}\Bigr]  & = 2(p-1) \bar{S}^{p,a}_{\bar{m},m}\\  \Bigl[   L_n,\bar S^{p,a}_{\bar{m},m}\Bigr]  &= (-)^n (n(p-1)-m)\bar S^{p,a}_{\bar{m},m+n} ,\hspace{26pt}~~~~\Bigl[  \bar L_{\bar{n}},\bar S^{p,a}_{\bar{m},m}\Bigr]  = (\bar{n}(1-p)-\bar{m})\bar S^{p,a}_{\bar{m}+\bar{n},m}.
        \label{eqn: S-bar SO(4,2) action}
    \end{split}
\end{equation}
Now consider the 2D OPE defining  $\ms^{1,a}$ as an operator on the celestial sphere  by \eqref{sdef}.   Take  $w\sim \phi-i\e$ a small distance $\e=\psi-{\pi \over 2}$ below  the real axis.\footnote{From the \ads\ perspective this amounts to moving the operator slightly inside the boundary.}   Using equation (4.3) of \cite{He:2015zea} we find the OPE\footnote{Double soft limits of differently-polarized  operators can have OPE ambiguities \cite{He:2015zea}. Here  both operators have the same (linear) polarization so there is no ambiguity in the OPE.}
\bea \ms^{1,a}(w_1, \bw_1) \ms^{1,b}(w_2, \bw_2)&\sim & \frac{-i f^{abc}}{2} \bigg({i\over \phi_{12}+i\e_{12} }-{i \over \phi_{12}-i\e_{12} }  \bigg)\ms^{1,c}(w_2, \bw_2)\cr 
&\sim&   -\pi i f^{abc}\hspace{2pt} \ms^{1,c}(\phi_2) \hspace{2pt} {\rm sgn}(\e_{12})\delta(\phi_{12}).\eea
The 2D euclidean radial-ordered  commutator  is then in modes
\bea\label{tmn}\!\! [\ms^{1,a}_m, \ms^{1,b}_n]&=& \int_0^{2\pi} \frac{d\phi_1}{2\pi}\frac{d\phi_2}{2\pi} e^{im\phi_1+in\phi_2} \lim_{\e_k\to 0}\Bigl[ \ms^{1,a}(\phi_1, \e_1) \ms^{1,b}(\phi_2, \e_2)|_{\e_1>\e_2}- \ms^{1,a}(\phi_1, \e_1) \ms^{1,b}(\phi_2, \e_2)|_{\e_1<\e_2} \Bigr]\cr
&=&-if^{abc}\ms^{1,c}_{m+n}. \label{eqn: p = 1 T algebra} \eea
This is the leading $p=1$ $S$-algebra. 

\section{CFT$_3$ light transforms}
\label{sec: CFT3 LR operators}
In this section we identify $T^{1,a}$ as a conserved current light transform in the CFT$_3$ boundary dual to quantum gravity in \ads. 

The boundary condition \eqref{ubc} sets the tangential components $F^a_{ij}|_{\psi={\pi \over 2}}=0 $ , where $i,j=t,\theta, \phi$. According to the \ads /CFT$_3$ dictionary, in the metric \eqref{adsmt},
\be ds^2_{AdS_4}={-dt^2 +d\theta^2\over \sin^2 \theta \cos^2\psi}+\sec^2\psi d\psi^2+\tan^2\psi d\phi^2, ~~\psi<{\pi  \over 2}\ee the rescaled normal components \footnote{Here  $\rj^a_i = -g_{YM}^{-2} F^a_{i\psi}|_{\psi={\pi \over 2}}$ with $g_{YM}^2=1$ so that the commutator of the global color charge is $\left[Q^a,Q^b\right]=-i f^{abc}Q^c$.}
\be \rj^a_i=-\sec^2 \psi n^\mu F^a_{i\mu}|_{\psi={\pi \over 2}}= - F^a_{i\psi}|_{\psi={\pi \over 2}} \ee comprise a dimension 2 conserved global symmetry current in the boundary CFT$_3$. In terms of the null coordinates \be t^\pm=\half(t\pm \theta)\ee we find that 
\be T^{1,a}(\phi)= 2\pi \int_0^{\pi} dt^+  \rj_+^a(t^+,t^-=0,\phi).\ee
Note  that in our conventions the induced metric on the boundary of \ads\ is a (divergent) constant times 
\be ds^{b2}_3= {-dt^2+d\theta^2 \over \sin^2\theta}+d\phi^2.\ee 
We wish to conformally map this to the standard metric on EC$^3$. See \cite{Hofman:2008ar, Cordova:2018ygx,Kologlu:2019bco,Kologlu:2019mfz,Karateev:2017jgd} for  the transformation properties of light ray operators.
Defining the Weyl factor
\be \Omega_{EC^3} =\sin \theta, \ee 
we obtain the  round metric on EC$^3$  
\be d s^2_3\to \Omega^2_{EC^3}ds^{b2}_3=-dt^2+d\theta^2 + \sin^2\theta d\phi^2, \ee while 
\be  \rj^a_i \to {1 \over \sin \theta}  \rj^a_i . \ee
One finds in the rescaled frame
\be T^{1,a}(\phi)= 2\pi \int_0^{\pi} {dt^+  \sin t^+} \rj_+^a(t^+,0,\phi)\equiv 2\pi \cl^a(\phi).\ee 
Here we have identified $\cl^a$ as the standard conserved current light transform    on  a null geodesic beginning at the south pole $\theta=0$ and ending at the north pole $\theta=\pi$ in EC$_3$, with   $\phi$ labeling  the polar angle.\footnote{These geodesic segments are particularly natural in the time-periodic CFTs considered in \cite{Melton:2025jee}, where they comprise half-orbits of any closed null geodesic.}  

The  3D commutators of $\cl^a(\phi)$   were computed in 
\cite{Cordova:2018ygx}\footnote{One may also  map the light transform to the  M$^3$ conformal frame primarily used in \cite{Cordova:2018ygx}. EC$^3$ is tessellated by M$^3$ diamonds. Conformally mapping one of these diamonds to the flat metric  $ds^2=-dy^+ dy^- + dy^2$, while choosing the beginning and ends of the geodesic to lie on $\scri^-$ and $\scri^+$ at $y^-=0$, one finds
\be  T^{1,a}(y) =  2\pi  \int_{-\infty}^{\infty} dy^+ J^{a}_+(y^+, y^- = 0,y)\equiv 2\pi \cl^a(y),~~~~\Bigl[ \cl^a(y_1),  \cl^b(y_2)\Bigr]=- if^{abc}\d(y_{12})\cl^c(y_2).\ee }\be \Bigl[ \cl^a(\phi_1),  \cl^b(\phi_2)\Bigr]=- if^{abc}\d(\phi_{12})\cl^c(\phi_2).\ee
This CFT$_3$ result agrees with the bulk result \eqref{tmn}. The  structure of the singularities as the light ray operators approach one another from different directions in \ec\ near $\p$\ads,  implies that the 2D Euclidean and 3D Lorentzian commutators coincide. In modes
\be \Bigl[ \cl^a_m,  \cl^b_n\Bigr]=- {i \over 2\pi} f^{abc}\cl^c_{m+n}.\ee 
Hence modes of the $\rj_i^a$ light transforms on EC$^3$ generate the leading $p=1$ $S$-algebra.

\section{Conformal tower of light ray operators }

The light ray operators $\cl^a_m$ do not form a complete multiplet under the boundary conformal group $SO(3,2)$. This group  are generated by the 10 $SO(4,2)$ generators which preserve $\p$\ads :\be  D;~~~~ L_n-\bar L_{-n};~~~~ K_{\half,-\half}, K_{-\half,\half},(K_{\half,\half}+K_{-\half,-\half});~~~~
P_{\half,-\half}, P_{-\half,\half},(P_{\half,\half}+P_{-\half,-\half}).\ee
The full  multiplet can be constructed by action with these $SO(3,2)$
generators. An efficient way to do this is to go back to the expression
\be \cl^{1,a}_{0,m}\equiv \cl^{a}_{m}=\frac{1}{4\pi}\left(S^{1,a}_{0,m}+ \bar S^{1,a}_{-\bar{m},0}\right).\ee
More well-defined CFT$_3$ operators may be obtained by starting with these and commuting  with any of the $SO(3,2)$ generators.\footnote{ From the bulk point of view, the corresponding modes are guaranteed to preserve the \ads\ boundary conditions  \eqref{ubc}.} Explicit expressions for the resulting  modes are readily obtained by using the actions of $SO(3,2)$ on $S$ and $\bar S$ given in \eqref{eq:[K,s]}-\eqref{l commutator} and \eqref{eqn: S-bar SO(4,2) action}. The most general  mode that can be obtained in this way is\footnote{In the Minkowskian analysis, the indices $p,\bar m,n$ denote  eigenvalues of the three $SO(4,2)$ Cartan generators $D, \bar L_0,  L_0$. In contrast $SO(3,2)$ has only the two Cartan generators $D$ and $L_0-\bar L_0$, and so the representations here contains $2p-1$ elements on each site of the weight lattice.} 
\be \cl^{p,a}_{\bar{n},m}\equiv \frac{1}{4\pi}\left(S^{p,a}_{\bar{n},m} + \bar{S}^{p,a}_{-\bar{m},-n}\right). \label{eqn: general light ray op definition} \ee
We show in Appendix \ref{appendix: lightray descendants} that $\cl^{p,a}_{\bar{m},m}$ comprise a complete and independent set of $SO(3,2)$ descendants of $\cl^{1,a}_{m}$. 

The algebra for these modes can be derived by applying  the Jacobi identity to their definition as $SO(3,2)$ descendants. One finds  
\be \Bigl[ \cl^{p,a}_{\bar m,m},\cl^{q,a}_{\bar n ,n}\Bigr]=-\frac{i}{2\pi} f^{abc}\cl^{p+q-1,c}_{\bar m+\bar n,m+n},\ee
which is of course the $S$-algebra. 

In conclusion, the light transforms  $\cl^a$ of a nonabelian conserved CFT$_3$ global symmetry current  $\rj_i^a$ along  null geodesics  beginning  and ending  at a pair of antipodal points in EC$^3$, together with their  $SO(3,2)$ conformal descendants,  form an infinite-dimensional $SO(3,2)$ multiplet whose commutators generate an $S$-algebra. 

\section*{Acknowledgments}

The authors would like to thank Nima Arkani-Hamed, Matthew Heydeman, Simon Heuveline, Mina Himwich, Daniel Jafferis, Lionel Mason, Monica Pate, Atul Sharma, David Skinner, Mark Spradlin, Anastasia Volovich and Xi Yin for sharing their expertise over many helpful conversations. This work is supported by DOE grant de-sc/0007870 and the Simons Collaboration on Celestial Holography.

\appendix

\section{CFT$_3$ light ray descendants}
\label{appendix: lightray descendants}

In Section \ref{sec: leading soft operator in AdS}, we noticed that $S^{1,a}(w)$ and $\bar{S}^{1,a}(\bar{w})$ individually do not satisfy the AdS$_4$ boundary conditions. Consider  the following two linear combinations 
\begin{equation}
    \begin{split}
        T^{1,a}(w,\bar{w}) &\equiv \frac{1}{2 i}\left(S^{1,a}(w) - \bar{S}^{1,a}(\bar w)\right)  ~ ~ \longrightarrow ~ ~ T^{1,a}_{0,m} = \frac{1}{2}\left(S^{1,a}_{0,m} + \bar{S}^{1,a}_{-\bar{m},0}\right)\\
        \widetilde{T}^{1,a}(w,\bar w) &\equiv \frac{1}{2 i}\left(S^{1,a}(w) + \bar{S}^{1,a}(\bar w)\right) ~ ~ \longrightarrow ~ ~ \widetilde{T}^{1,a}_{0,m} = \frac{1}{2}\left(S^{1,a}_{0,m} - \bar{S}^{1,a}_{-\bar{m},0}\right)
    \end{split}
\end{equation}
The $T^{1,a}_{0,m}$ modes are permitted by the boundary conditions while the $\widetilde{T}^{1,a}_{0,m}$ modes are projected out. These modes are not closed under the action of the boundary $SO(3,2)$, so we define
\begin{equation}
    T^{p,a}_{\bar{n},m} = \frac{1}{2}\left(S^{p,a}_{\bar{n},m} + \bar{S}^{p,a}_{-\bar{m},-n}\right) \hspace{40pt} \widetilde{T}^{p,a}_{\bar{n},m} = \frac{1}{2}\left(S^{p,a}_{\bar{n},m} - \bar{S}^{p,a}_{-\bar{m},-n}\right)
\end{equation}
These modes are all linearly independent because the $S^{p,a}_{\bar{m},m}$ modes which they are comprised of are linearly independent.

In Section \ref{sec: S algebra from conf descendents}, we  built all the $S^{p,a}_{\bar{m},m}$ modes using $SO(4,2)$ bulk conformal transformations and the leading soft algebra among the $S^{1,a}_{0,m}$ modes. Now, we are able to build a full set of $T^{p,a}_{\bar{m},m}$ modes using only $SO(3,2)$ boundary conformal transformations \eqref{sott} and the leading soft algebra among the $T^{1,a}_{0,m}$ modes \eqref{eqn: p = 1 T algebra}. One  can raise the $p$ index to any value with
\begin{equation}
    \Bigl[K_{\pm \frac{1}{2},\mp\frac{1}{2}},T^{p,a}_{\bar{m},m}\Bigr] = (1-p\pm m) T^{p+1/2,a}_{\bar{m}\pm \half,m\mp \half} .\\
    \label{eqn: K action on T}
\end{equation}
$m$  and $\bar m$ can then be set to any value by the combined  action of $T^{1,a}_{0,m}$ and $L_n-\bar L_{-n}$.

An identical equation holds where we replace $T^{p,a}_{\bar{m},m} \leftrightarrow \widetilde{T}^{p,a}_{\bar{m},m}$. Because $\widetilde{T}^{1,a}_{0,m}$ are killed by the boundary condition, all $\widetilde{T}^{p,a}_{\bar{m},m}$ must also be killed. It is sensible that our Dirichlet boundary conditions project out precisely half of the degrees of freedom.

Using the techniques of Section \ref{sec: S algebra from conf descendents}, it is straightforward to show that the modes that we have built satisfy the S-algebra as well
\begin{equation}
    [T^{p,a}_{\bar{m},m},T^{q,b}_{\bar{n},n}] = -i f^{abc}T^{p + q - 1,c}_{\bar{m} + \bar{n},m + n}.
\end{equation}
We have demonstrated that the $T^{p,a}_{\bar{m},m}$ modes defined above form a complete set of objects which are closed under boundary conformal transformations and commutators among themselves. 

In Section \ref{sec: CFT3 LR operators}, we argued that $T^{1,a}$ may be reinterpreted as a light-transformed current  in CFT$_3$, 
\begin{equation}
    T^{1,a}(\phi)= 2\pi\int_0^{\pi} dt^+ \hspace{2pt} \sin t^+ \hspace{2pt} \rj_+^a(t,\phi,\theta)\equiv 2\pi \cl^a(\phi).
\end{equation}
Just as we have built a closed algebra of $T^{p,a}_{\bar{m},m}$ modes by acting on $T^{1,a}_{0,m}$ with elements of the boundary conformal group, we may now act directly on the CFT$_3$ light ray operators with such boundary conformal transformations. In this way, we build a family of light ray operators $\cl^{p,a}_{\bar{m},m}$. These are literally the same object as $T^{p,a}_{\bar{m},m}$, just built out of the CFT$_3$ data. Therefore, they must satisfy equation \eqref{eqn: K action on T} and obey the same algebra
\begin{equation}
\Bigl[\cl^{p,a}_{\bar{m},m},\cl^{q,b}_{\bar{n},n}\Bigr] = -\frac{i}{2\pi}f^{abc} \cl^{p + q - 1,c}_{\bar{m} + \bar{n},m + n}.
\end{equation}

\bibliographystyle{JHEP}
\bibliography{refs.bib}

\providecommand{\href}[2]{#2}\begingroup\raggedright\begin{thebibliography}{10}

\bibitem{Guevara:2021abz}
A.~Guevara, E.~Himwich, M.~Pate, and A.~Strominger, {\it {Holographic symmetry algebras for gauge theory and gravity}},  {\em JHEP} {\bf 11} (2021) 152, [\href{http://arxiv.org/abs/2103.03961}{{\tt arXiv:2103.03961}}].

\bibitem{Strominger:2021mtt}
A.~Strominger, {\it {$w_{1+\infty}$ Algebra and the Celestial Sphere: Infinite Towers of Soft Graviton, Photon, and Gluon Symmetries}},  {\em Phys. Rev. Lett.} {\bf 127} (2021), no.~22 221601, [\href{http://arxiv.org/abs/2105.14346}{{\tt arXiv:2105.14346}}].

\bibitem{Penrose:1976jq}
R.~Penrose, {\it {The Nonlinear Graviton}},  {\em Gen. Rel. Grav.} {\bf 7} (1976) 171--176.

\bibitem{Ward:1977ta}
R.~S. Ward, {\it {On Selfdual gauge fields}},  {\em Phys. Lett. A} {\bf 61} (1977) 81--82.

\bibitem{Costello:2022wso}
K.~Costello and N.~M. Paquette, {\it {Celestial holography meets twisted holography: 4d amplitudes from chiral correlators}},  {\em JHEP} {\bf 10} (2022) 193, [\href{http://arxiv.org/abs/2201.02595}{{\tt arXiv:2201.02595}}].

\bibitem{Costello:2022jpg}
K.~Costello, N.~M. Paquette, and A.~Sharma, {\it {Top-Down Holography in an Asymptotically Flat Spacetime}},  {\em Phys. Rev. Lett.} {\bf 130} (2023), no.~6 061602, [\href{http://arxiv.org/abs/2208.14233}{{\tt arXiv:2208.14233}}].

\bibitem{Adamo:2021lrv}
T.~Adamo, L.~Mason, and A.~Sharma, {\it {Celestial $w_{1+\infty}$ Symmetries from Twistor Space}},  {\em SIGMA} {\bf 18} (2022) 016, [\href{http://arxiv.org/abs/2110.06066}{{\tt arXiv:2110.06066}}].

\bibitem{Kmec:2025ftx}
A.~Kmec, L.~Mason, R.~Ruzziconi, and A.~Sharma, {\it {S-algebra in gauge theory: twistor, spacetime and holographic perspectives}},  {\em Class. Quant. Grav.} {\bf 42} (2025), no.~19 195008, [\href{http://arxiv.org/abs/2506.01888}{{\tt arXiv:2506.01888}}].

\bibitem{Brown:1986nw}
J.~D. Brown and M.~Henneaux, {\it {Central Charges in the Canonical Realization of Asymptotic Symmetries: An Example from Three-Dimensional Gravity}},  {\em Commun. Math. Phys.} {\bf 104} (1986) 207--226.

\bibitem{Maldacena:1997re}
J.~M. Maldacena, {\it {The Large $N$ limit of superconformal field theories and supergravity}},  {\em Adv. Theor. Math. Phys.} {\bf 2} (1998) 231--252, [\href{http://arxiv.org/abs/hep-th/9711200}{{\tt hep-th/9711200}}].

\bibitem{Aharony:1999ti}
O.~Aharony, S.~S. Gubser, J.~M. Maldacena, H.~Ooguri, and Y.~Oz, {\it {Large N field theories, string theory and gravity}},  {\em Phys. Rept.} {\bf 323} (2000) 183--386, [\href{http://arxiv.org/abs/hep-th/9905111}{{\tt hep-th/9905111}}].

\bibitem{tHooft:1993dmi}
G.~'t~Hooft, {\it {Dimensional reduction in quantum gravity}},  {\em Conf. Proc. C} {\bf 930308} (1993) 284--296, [\href{http://arxiv.org/abs/gr-qc/9310026}{{\tt gr-qc/9310026}}].

\bibitem{Susskind:1994vu}
L.~Susskind, {\it {The World as a hologram}},  {\em J. Math. Phys.} {\bf 36} (1995) 6377--6396, [\href{http://arxiv.org/abs/hep-th/9409089}{{\tt hep-th/9409089}}].

\bibitem{deGioia:2023cbd}
L.~P. de~Gioia and A.-M. Raclariu, {\it {Celestial sector in CFT: Conformally soft symmetries}},  {\em SciPost Phys.} {\bf 17} (2024), no.~1 002, [\href{http://arxiv.org/abs/2303.10037}{{\tt arXiv:2303.10037}}].

\bibitem{deGioia:2024yne}
L.~P. de~Gioia and A.-M. Raclariu, {\it {Celestial amplitudes from conformal correlators with bulk-point kinematics}},  \href{http://arxiv.org/abs/2405.07972}{{\tt arXiv:2405.07972}}.

\bibitem{deGioia:2025mwt}
L.~P. de~Gioia and A.-M. Raclariu, {\it {Infinite towers of 2d symmetry algebras from Carrollian limit of 3d CFT}},  \href{http://arxiv.org/abs/2508.19981}{{\tt arXiv:2508.19981}}.

\bibitem{Ward:cc}
R.~S. Ward, {\it Self-dual space-times with cosmological constant},  {\em Communications in Mathematical Physics} {\bf 78} (1980), no.~1 1--17.

\bibitem{Taylor:2023ajd}
T.~R. Taylor and B.~Zhu, {\it {w1+{\ensuremath{\infty}} Algebra with a Cosmological Constant and the Celestial Sphere}},  {\em Phys. Rev. Lett.} {\bf 132} (2024), no.~22 221602, [\href{http://arxiv.org/abs/2312.00876}{{\tt arXiv:2312.00876}}].

\bibitem{Bittleston:2024rqe}
R.~Bittleston, G.~Bogna, S.~Heuveline, A.~Kmec, L.~Mason, and D.~Skinner, {\it {On AdS$_{4}$ deformations of celestial symmetries}},  {\em JHEP} {\bf 07} (2024) 010, [\href{http://arxiv.org/abs/2403.18011}{{\tt arXiv:2403.18011}}].

\bibitem{Aharony:2008ug}
O.~Aharony, O.~Bergman, D.~L. Jafferis, and J.~Maldacena, {\it {N=6 superconformal Chern-Simons-matter theories, M2-branes and their gravity duals}},  {\em JHEP} {\bf 10} (2008) 091, [\href{http://arxiv.org/abs/0806.1218}{{\tt arXiv:0806.1218}}].

\bibitem{Hofman:2008ar}
D.~M. Hofman and J.~Maldacena, {\it {Conformal collider physics: Energy and charge correlations}},  {\em JHEP} {\bf 05} (2008) 012, [\href{http://arxiv.org/abs/0803.1467}{{\tt arXiv:0803.1467}}].

\bibitem{Casini:2017roe}
H.~Casini, E.~Teste, and G.~Torroba, {\it {Modular Hamiltonians on the null plane and the Markov property of the vacuum state}},  {\em J. Phys. A} {\bf 50} (2017), no.~36 364001, [\href{http://arxiv.org/abs/1703.10656}{{\tt arXiv:1703.10656}}].

\bibitem{Karateev:2017jgd}
D.~Karateev, P.~Kravchuk, and D.~Simmons-Duffin, {\it {Weight Shifting Operators and Conformal Blocks}},  {\em JHEP} {\bf 02} (2018) 081, [\href{http://arxiv.org/abs/1706.07813}{{\tt arXiv:1706.07813}}].

\bibitem{Cordova:2018ygx}
C.~C{\'o}rdova and S.-H. Shao, {\it {Light-ray Operators and the BMS Algebra}},  {\em Phys. Rev. D} {\bf 98} (2018), no.~12 125015, [\href{http://arxiv.org/abs/1810.05706}{{\tt arXiv:1810.05706}}].

\bibitem{Kravchuk:2018htv}
P.~Kravchuk and D.~Simmons-Duffin, {\it {Light-ray operators in conformal field theory}},  {\em JHEP} {\bf 11} (2018) 102, [\href{http://arxiv.org/abs/1805.00098}{{\tt arXiv:1805.00098}}].

\bibitem{Kologlu:2019mfz}
M.~Kologlu, P.~Kravchuk, D.~Simmons-Duffin, and A.~Zhiboedov, {\it {The light-ray OPE and conformal colliders}},  {\em JHEP} {\bf 01} (2021) 128, [\href{http://arxiv.org/abs/1905.01311}{{\tt arXiv:1905.01311}}].

\bibitem{Kologlu:2019bco}
M.~Kologlu, P.~Kravchuk, D.~Simmons-Duffin, and A.~Zhiboedov, {\it {Shocks, Superconvergence, and a Stringy Equivalence Principle}},  {\em JHEP} {\bf 11} (2020) 096, [\href{http://arxiv.org/abs/1904.05905}{{\tt arXiv:1904.05905}}].

\bibitem{Belin:2020lsr}
A.~Belin, D.~M. Hofman, G.~Mathys, and M.~T. Walters, {\it {On the stress tensor light-ray operator algebra}},  {\em JHEP} {\bf 05} (2021) 033, [\href{http://arxiv.org/abs/2011.13862}{{\tt arXiv:2011.13862}}].

\bibitem{Huang:2019fog}
K.-W. Huang, {\it {Stress-tensor commutators in conformal field theories near the lightcone}},  {\em Phys. Rev. D} {\bf 100} (2019), no.~6 061701, [\href{http://arxiv.org/abs/1907.00599}{{\tt arXiv:1907.00599}}].

\bibitem{Huang:2020ycs}
K.-W. Huang, {\it {Lightcone Commutator and Stress-Tensor Exchange in $d>2$ CFTs}},  {\em Phys. Rev. D} {\bf 102} (2020), no.~2 021701, [\href{http://arxiv.org/abs/2002.00110}{{\tt arXiv:2002.00110}}].

\bibitem{Kabat:2021akg}
D.~Kabat, G.~Lifschytz, P.~Nguyen, and D.~Sarkar, {\it {Light-ray moments as endpoint contributions to modular Hamiltonians}},  {\em JHEP} {\bf 09} (2021) 074, [\href{http://arxiv.org/abs/2103.08636}{{\tt arXiv:2103.08636}}].

\bibitem{Mishra_2018}
R.~K. Mishra and R.~Sundrum, {\it Asymptotic symmetries, holography and topological hair},  {\em Journal of High Energy Physics} {\bf 2018} (Jan., 2018).

\bibitem{Mishra_2019}
R.~K. Mishra, A.~Mohd, and R.~Sundrum, {\it Ads asymptotic symmetries from cft mirrors},  {\em Journal of High Energy Physics} {\bf 2019} (Mar., 2019).

\bibitem{He:2019ywq}
T.~He and P.~Mitra, {\it {New magnetic symmetries in $(d + 2)$-dimensional QED}},  {\em JHEP} {\bf 01} (2021) 122, [\href{http://arxiv.org/abs/1907.02808}{{\tt arXiv:1907.02808}}].

\bibitem{Donnay:2020fof}
L.~Donnay, G.~Giribet, and F.~Rosso, {\it {Quantum BMS transformations in conformally flat space-times and holography}},  {\em JHEP} {\bf 12} (2020) 102, [\href{http://arxiv.org/abs/2008.05483}{{\tt arXiv:2008.05483}}].

\bibitem{Gonzo:2020xza}
R.~Gonzo and A.~Pokraka, {\it {Light-ray operators, detectors and gravitational event shapes}},  {\em JHEP} {\bf 05} (2021) 015, [\href{http://arxiv.org/abs/2012.01406}{{\tt arXiv:2012.01406}}].

\bibitem{Hu:2022txx}
Y.~Hu and S.~Pasterski, {\it {Celestial conformal colliders}},  {\em JHEP} {\bf 02} (2023) 243, [\href{http://arxiv.org/abs/2211.14287}{{\tt arXiv:2211.14287}}].

\bibitem{Donnay:2022aba}
L.~Donnay, A.~Fiorucci, Y.~Herfray, and R.~Ruzziconi, {\it {Carrollian Perspective on Celestial Holography}},  {\em Phys. Rev. Lett.} {\bf 129} (2022), no.~7 071602, [\href{http://arxiv.org/abs/2202.04702}{{\tt arXiv:2202.04702}}].

\bibitem{Donnay:2022wvx}
L.~Donnay, A.~Fiorucci, Y.~Herfray, and R.~Ruzziconi, {\it {Bridging Carrollian and celestial holography}},  {\em Phys. Rev. D} {\bf 107} (2023), no.~12 126027, [\href{http://arxiv.org/abs/2212.12553}{{\tt arXiv:2212.12553}}].

\bibitem{Hu:2023geb}
Y.~Hu and S.~Pasterski, {\it {Detector operators for celestial symmetries}},  {\em JHEP} {\bf 12} (2023) 035, [\href{http://arxiv.org/abs/2307.16801}{{\tt arXiv:2307.16801}}].

\bibitem{Gonzalez:2025ene}
H.~A. Gonz{\'a}lez and J.~Salzer, {\it {Energy Detectors and Asymptotic Symmetries}},  \href{http://arxiv.org/abs/2510.27348}{{\tt arXiv:2510.27348}}.

\bibitem{Moult:2025njc}
I.~Moult, S.~A. Narayanan, and S.~Pasterski, {\it {Memory Correlators and Ward Identities in the 'in-in' Formalism}},  \href{http://arxiv.org/abs/2512.02825}{{\tt arXiv:2512.02825}}.

\bibitem{Himwich:2025ekg}
E.~Himwich and M.~Pate, {\it {Light-ray Operators and the ${\rm w}_{1+\infty}$ Algebra}},  \href{http://arxiv.org/abs/2512.18973}{{\tt arXiv:2512.18973}}.

\bibitem{Chalmers:1996rq}
G.~Chalmers and W.~Siegel, {\it {The Selfdual sector of QCD amplitudes}},  {\em Phys. Rev. D} {\bf 54} (1996) 7628--7633, [\href{http://arxiv.org/abs/hep-th/9606061}{{\tt hep-th/9606061}}].

\bibitem{Ball:2021tmb}
A.~Ball, S.~A. Narayanan, J.~Salzer, and A.~Strominger, {\it {Perturbatively exact w$_{1+\infty}$ asymptotic symmetry of quantum self-dual gravity}},  {\em JHEP} {\bf 01} (2022) 114, [\href{http://arxiv.org/abs/2111.10392}{{\tt arXiv:2111.10392}}].

\bibitem{Costello:2018zrm}
K.~Costello and D.~Gaiotto, {\it {Twisted holography}},  {\em JHEP} {\bf 01} (2025) 087, [\href{http://arxiv.org/abs/1812.09257}{{\tt arXiv:1812.09257}}].

\bibitem{Costello:2023vyy}
K.~J. Costello, {\it {Bootstrapping two-loop QCD amplitudes}},  {\em JHEP} {\bf 08} (2025) 011, [\href{http://arxiv.org/abs/2302.00770}{{\tt arXiv:2302.00770}}].

\bibitem{Bittleston:2024efo}
R.~Bittleston, K.~Costello, and K.~Zeng, {\it {Self-Dual Gauge Theory from the Top Down}},  \href{http://arxiv.org/abs/2412.02680}{{\tt arXiv:2412.02680}}.

\bibitem{Banks:1981nn}
T.~Banks and A.~Zaks, {\it {On the Phase Structure of Vector-Like Gauge Theories with Massless Fermions}},  {\em Nucl. Phys. B} {\bf 196} (1982) 189--204.

\bibitem{Seiberg:1994pq}
N.~Seiberg, {\it {Electric - magnetic duality in supersymmetric nonAbelian gauge theories}},  {\em Nucl. Phys. B} {\bf 435} (1995) 129--146, [\href{http://arxiv.org/abs/hep-th/9411149}{{\tt hep-th/9411149}}].

\bibitem{Strominger:2013lka}
A.~Strominger, {\it {Asymptotic Symmetries of Yang-Mills Theory}},  {\em JHEP} {\bf 07} (2014) 151, [\href{http://arxiv.org/abs/1308.0589}{{\tt arXiv:1308.0589}}].

\bibitem{He:2015zea}
T.~He, P.~Mitra, and A.~Strominger, {\it {2D Kac-Moody Symmetry of 4D Yang-Mills Theory}},  {\em JHEP} {\bf 10} (2016) 137, [\href{http://arxiv.org/abs/1503.02663}{{\tt arXiv:1503.02663}}].

\bibitem{Strominger:2017zoo}
A.~Strominger, {\em {Lectures on the Infrared Structure of Gravity and Gauge Theory}}.
\newblock Princeton University Press, 2018.

\bibitem{Larkoski:2014hta}
A.~J. Larkoski, {\it {Conformal Invariance of the Subleading Soft Theorem in Gauge Theory}},  {\em Phys. Rev. D} {\bf 90} (2014), no.~8 087701, [\href{http://arxiv.org/abs/1405.2346}{{\tt arXiv:1405.2346}}].

\bibitem{Chen:2023tvj}
H.~Z. Chen, R.~C. Myers, and A.-M. Raclariu, {\it {Entanglement, soft modes, and celestial holography}},  {\em Phys. Rev. D} {\bf 109} (2024), no.~12 L121702, [\href{http://arxiv.org/abs/2308.12341}{{\tt arXiv:2308.12341}}].

\bibitem{Pasterski:2016qvg}
S.~Pasterski, S.-H. Shao, and A.~Strominger, {\it {Flat Space Amplitudes and Conformal Symmetry of the Celestial Sphere}},  {\em Phys. Rev. D} {\bf 96} (2017), no.~6 065026, [\href{http://arxiv.org/abs/1701.00049}{{\tt arXiv:1701.00049}}].

\bibitem{Pasterski_Shao_2017}
S.~Pasterski and S.-H. Shao, {\it Conformal basis for flat space amplitudes},  {\em Physical Review D} {\bf 96} (Sept., 2017).

\bibitem{Jain:2024bza}
S.~Jain, D.~K. S, and E.~Skvortsov, {\it {Hidden sectors of Chern-Simons matter theories and exact holography}},  {\em Phys. Rev. D} {\bf 111} (2025), no.~10 106017, [\href{http://arxiv.org/abs/2405.00773}{{\tt arXiv:2405.00773}}].

\bibitem{Aharony:2024nqs}
O.~Aharony, R.~R. Kalloor, and T.~Kukolj, {\it {A chiral limit for Chern-Simons-matter theories}},  {\em JHEP} {\bf 10} (2024) 051, [\href{http://arxiv.org/abs/2405.01647}{{\tt arXiv:2405.01647}}].

\bibitem{Melton:2025jee}
W.~Melton and A.~Strominger, {\it {Conformal Field Theory with Periodic Time}},  \href{http://arxiv.org/abs/2512.09089}{{\tt arXiv:2512.09089}}.

\end{thebibliography}\endgroup


\end{document}